\documentclass[twoside]{article}

\usepackage{fleqn,espcrc2}
\usepackage{epsfig}
\usepackage[figuresright]{rotating}

\newcommand{\AmS}{{\protect\the\textfont2
  A\kern-.1667em\lower.5ex\hbox{M}\kern-.125emS}}

\newcommand{\vp}{\mbox{\boldmath$p$}}
\newcommand{\vk}{\mbox{\boldmath$k$}}
\newcommand{\vq}{\mbox{\boldmath$q$}}
\newcommand{\vg}{\mbox{\boldmath$\gamma$}}
\def\beq{\begin{equation}}
\def\eeq{\end{equation}}
\def\bea{\begin{eqnarray}}
\def\eea{\end{eqnarray}}

\title{Van Hove Singularities in the Quark-Gluon Plasma}

\author{Markus H. Thoma\address{Theory Division, CERN, CH-1211 Geneva 23, 
Switzerland}%
\thanks{Heisenberg Fellow}}

\begin{document}

\begin{abstract}
General arguments as well as different approximations for the in-medium
quark propagator in a quark-gluon plasma lead to quark dispersion relations
that exhibit a minimum in one branch (plasmino). This minimum causes 
Van Hove singularities in the dilepton 
production rate and mesonic correlators, which might have observable
consequences. 
\vspace{1pc}
\end{abstract}

\maketitle



\section{Introduction}

In 1953 Van Hove \cite{Van53} discussed singularities in 
the density of states, so-called Van Hove singularities,
in solid state physics. The density of states of a system, given by
\beq
g(\omega)=\sum_n \int \frac{{\rm d}^3k}{(2\pi )^3}\> \delta 
(\omega -\omega_n(\vk))
\label{eq1}
\eeq
with the energy eigenstates $\omega_n({\vk})$, can be expressed
by the surface integral \cite{Ash76}
\beq
g(\omega)=\sum_n \int \frac{{\rm d}S}{(2\pi )^3}\> 
\frac{1}{|{\bf \nabla} \omega_n(\vk)|}.
\label{eq2}
\eeq
Here the quantity in the denominator $|{\bf \nabla} \omega_n({\vk})|$
can be identified with the group velocity. Due to symmetries in a crystal
the group velocity vanishes at certain momenta, resulting in a divergent
integrand in (\ref{eq2}). This divergence is integrable in 3 dimensions,
leading to a finite density of states. In lower dimensions, however,
Van Hove singularities appear. For example, a 2-dimensional electron
gas shows logarithmic singularities, which have been discussed in connection 
with high-$T_c$ superconductors \cite{Mar97}.

Here we want to discuss the role of Van Hove singularities in a quark-gluon 
plasma. We will argue that the in-medium quark dispersion relation
consists of two branches of which one has a minimum at some finite value of the
momentum. This leads to a vanishing group velocity for the collective
quark modes. Interesting quantities such as the production rate of low
mass lepton pairs and mesonic correlators depend inversely on this
group velocity. Therefore these quantities, which follow from self energy 
diagrams containing a quark loop, are affected by Van Hove singularities,
which might have observable consequences.

In the next two sections we will show the origin of Van Hove singularities 
using an effective quark propagator for the quark loop in the self energies
under consideration. First we will use the so-called hard thermal loop
(HTL) resummed quark propagator. After that we discuss the implications
of a quark propagator considering the presence of a non-vanishing 
gluon condensate in the deconfined phase as indicated by lattice
QCD calculations. In section 4 we will argue that the minimum in the
quark dispersion relation and thus the appearance of Van Hove singularities 
is a general property of massless fermions at finite temperature. Finally
we will discuss the chances and problems to observe Van Hove singularities
and to prove in this way the presence of deconfined collective quarks in
relativistic heavy ion collisions.

\section{Hard thermal loop approximation}

The HTL resummation technique has been developed by Braaten and Pisarski to
cure some serious problems of perturbation theory for gauge theories
at finite temperature \cite{Bra90}. Using only bare propagators and
vertices naive perturbation theory lead to gauge dependent and infrared
divergent results for physical quantities. A famous example is the 
damping rate of a long-wave, collective gluon mode in the QGP, which was found
to depend on the gauge choice and became even negative corresponding to a
plasma instability in covariant gauges \cite{Lop85}. Braaten and Pisarski
realized that this undesirable behaviour is due to the fact that higher
order loop diagrams contribute to lower order in the coupling constant at 
finite temperature. Distinguishing between a soft momentum scale $gT$,
where $g$ is the gauge coupling, and 
a hard scale of order $T$, Braaten and Pisarski isolated the dangerous 
diagrams. They are given by one-loop self energies and vertices, which contain only
hard loop momenta and are related to each other by Ward identities.
For example, in the case of the polarization tensor
in QED or QCD a result proportional to $g^2T^2$ was obtained\footnote{This
result has been found before in the high temperature approximation
\cite{Kli82} and much earlier already using the semiclassical Vlasov equation
\cite{Sil60}.}. Resumming these self energies and vertices within the
Dyson-Schwinger equation effective propagators and vertices can be 
constructed. These effective Green functions have to be used
if all legs of the Green function under consideration are soft. 
Otherwise bare propagators and vertices are sufficient. Physical quantities
calculated by applying this HTL improved perturbation theory, such as the
gluon damping rate \cite{Bra90a}, turn out to be gauge invariant. At the
same time medium effects such as Debye screening, leading to an improved
infrared behaviour, and Landau damping are included. For a review of the
HTL resummation method and its applications to the physics of the QGP see
Ref.\cite{Tho95}.

Let us now consider the HTL approximation for the quark propagator
in the QGP. Assuming that the temperature is much larger than the mass,
which holds at least for up and down quarks, we can neglect the quark masses.
Then the most general expression for the fermion self energy in the heat
bath can be written as \cite{Wel82} 
\beq
\Sigma(K)=-a(k_0,k) K^\mu \gamma_\mu - b(k_0,k)\gamma_0,
\label{eq3}
\eeq
using the notation $K\equiv (k_0, \vk)$, $k\equiv |\vk|$. 
The scalar quantities $a$ and $b$ are functions of the energy $k_0$ and
magnitude $k$ of the three momentum. They are given by traces over
the self energy
\bea
a(k_0,k) \hspace*{-0.2cm} & = & \hspace*{-0.2cm} 
\frac {1}{4k^2}\> \left [tr(K^\mu \gamma _\mu\> \Sigma )
- k_0\> tr(\gamma _0 \> \Sigma )\right ],\hspace*{-0.2cm} \label{eq4}\\
b(k_0,k) \hspace*{-0.2cm} & = & \hspace*{-0.2cm} 
\frac {1}{4k^2}\> \left [K^2\> tr(\gamma _0\> \Sigma )
- k_0\> tr (K^\mu \gamma _\mu \> \Sigma )\right ],
\nonumber
\eea
which read in the HTL approximation
\bea
tr(K^\mu \gamma _\mu \> \Sigma ) & = & 4\> m_q^2\; ,\nonumber \\
tr(\gamma _0\> \Sigma ) & = & 2\> m_q^2\> \frac {1}{k}\> \ln \frac
{k_0+k}{k_0-k}
\label{eq5}
\eea
with the effective thermal quark mass $m_q^2=g^2T^2/6$. Note that the
general ansatz (\ref{eq3}) is chirally invariant in spite of the occurrence
of an effective quark mass \cite{Wel82}. 
Furthermore the quark self energy has a
non-vanishing imaginary part below the light cone ($k_0^2<k^2$) which
can be related to Landau damping for spacelike quark momenta.

For massless quarks it is convenient to decompose the quark propagator 
into its helicity eigenstates 
($\hat{\vk}=\vk /k$) \cite{Bra90b}
\beq
S(K) = \frac{\gamma_0-\hat{\mbox{\boldmath$k$}}\cdot 
\mbox{\boldmath$\gamma$}}{2D_+(K)}
+\frac{\gamma_0+\hat{\mbox{\boldmath$k$}}\cdot 
\mbox{\boldmath$\gamma$}}{2D_-(K)}.
\label{eq6}
\eeq
The zeros $\omega_\pm(k)$ of
\beq
D_\pm = (1+a)\> (-k_0\pm k)-b,
\label{eq7}
\eeq
describe the dispersion relation of the particle excitation $q_+$ with energy
$\omega_+$ and of a mode $q_-$, called plasmino \cite{Bra90b}, with
energy $\omega_-$ and negative ratio of chirality to helicity. The
latter is a consequence of the medium, breaking the Lorentz invariance
of the vacuum. 

The HTL quark dispersion relations are shown in Fig.1. We observe that
both branches start at zero momentum at the same energy $\omega_\pm(0)=m_q$
and approach the bare dispersion $\omega =k$ at large momenta.
The plasmino branch, which is absent in the vacuum, has a minimum at
$k=0.408\, m_q$. The spectral strength of the plasmino decreases
exponentially for large momenta, indicating the purely collective nature of 
this mode. 

\begin{figure}
\vspace*{-11.5cm}
\hspace*{4.4cm}
\centerline{\psfig{figure=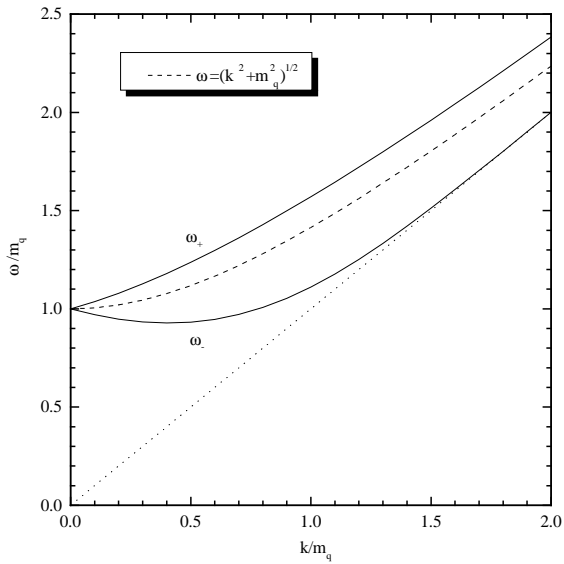,width=8.5cm}}
\vspace*{4.3cm}
\caption{Quark dispersion relation in HTL approximation. Also shown are
the free dispersion relations of massless (dotted line) and massive 
(dashed line) quarks.}
\end{figure}

The emission of thermal lepton pairs from the QGP proceeds via the decay of 
a virtual photon produced to lowest order by quark-antiquark annihilation.
For low invariant masses of the order $gT$ medium effects from the
collective quark modes have to be taken into account. The annihilation
amplitude is related to the imaginary part of the photon self energy
containing a quark loop by cutting rules \cite{Wel83}. According to
the rules of the HTL method one has to use effective propagators
and vertices as shown in Fig.2, if the photon energy and momentum
are soft. The imaginary part of this diagram comes either from the pole 
of the effective quark propagator, corresponding to the dispersion relations
of Fig.1, or from the the imaginary part of the HTL quark self energy
contained in the resummed propagator (cut contribution). Therefore we
have pole-pole, pole-cut, and cut-cut contributions to the
dilepton production rate which is proportional to the imaginary part
of the polarization tensor according to \cite{Gal91}
\beq
\frac{dN}{d^4xd^4p}=\frac{\alpha}{12\pi^4}\> \frac{1}{{\rm e}^{E/T}-1}\>
\frac{{\rm Im}\, {\Pi^\mu}_\mu (P)}{M^2}
\label{eq7a}
\eeq
with the QED fine-structure constant $\alpha $ and 
the invariant photon mass $M^2\equiv P^2$. The pole-cut and cut-cut
contributions involving external gluons, as can be seen by cutting the HTL 
quark self energy, lead to a smooth contribution to the dilepton rate
\cite{Bra90b}. The pole-pole term, on the other hand, 
will cause sharp structures as we will discuss below.

\begin{figure}
\centerline{\psfig{figure=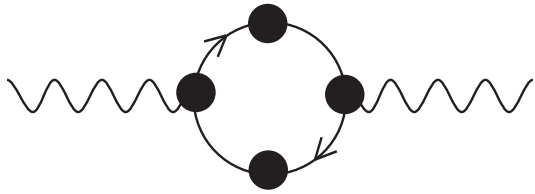,width=7cm}}
\vspace*{-0.5cm}
\caption{Photon self energy containing effective quark propagators and 
quark-photon vertices.}
\end{figure}

Besides the dilepton production rates also temporal correlators 
of mesons follow from the diagram of Fig.2 \cite{Kar00}. These
correlation functions of meson currents at the same space coordinate
but different points in Euclidean time have been measured on the lattice 
in the deconfined phase (see e.g. Ref.\cite{Boy94}). A clear deviation 
from the free correlator, which is determined using only bare quark propagators
and vertices in Fig.2 \cite{Flo94}, has been found in particular in the
pion channel \cite{Boy94}. 
This indicates the importance of quark interactions and
medium effects in the QGP at temperatures close to the critical. Using
HTL propagators and vertices as in Fig.2, medium effects, namely
effective quark masses and Landau damping, are taken into account. It is
interesting to study to what extend the HTL medium effects already explain
the lattice results.

The spectral functions of the temporal correlators are proportional
to the imaginary part of the quark loop diagram. The only difference 
compared to the photon self energy lies in the vertices describing the various
mesonic channels. For example, in the pion channel the bare vertex is 
proportional to $\gamma_5$. In the vector channel, on the other hand,
the bare vertex is proportional to $\gamma_\mu$ as
in the dilepton case corresponding to Vector Meson Dominance.

In contrast to photons and vector mesons it is sufficient to consider a 
bare meson-quark vertex in Fig.2 for pseudo-scalar mesons. This is due
to the fact that HTL vertices in the case of the Yukawa theory lead only
to higher order contributions and can be neglected therefore \cite{Tho95a}. 
Since HTL vertices are complicated functions of the energy and momentum,
we will discuss here only the spectral function of the pseudo-scalar
temporal correlator. Actually the consideration of the effective vertex
does not change the position of the Van Hove singularities, which are 
determined only by the minimum in the plasmino dispersion. Furthermore in 
the dilepton rate from the QGP the inclusion or neglect of the effective vertex
does also not alter the magnitude of the rate significantly. This is in 
contrast to the dilepton rate from pion annihilation, where the
consideration of the Ward identity reduced similar singularities in the
rate strongly \cite{Kor91}.

The spectral function of the pseudo-scalar correlator is given by 
\begin{equation}
\sigma_{ps}(\omega,\vp) = {1\over \pi}\> {\rm Im}\;  
\chi_{ps}(\omega,\vp), 
\label{eq8}
\end{equation} 
where the correlation function in momentum space contains the quark loop
according to
\bea
\chi_{ps}(\omega, \vp)\hspace*{-0.2cm} &=& \hspace*{-0.2cm} 2 N_c 
T \sum_n \int {{\rm d}^3k \over (2 \pi)^3} \;{\rm Tr}\biggl[ \gamma_5
S_{\rm HTL}(k_0,\vk) \nonumber \\
\hspace*{-0.2cm}&\times &\hspace*{-0.2cm}
\gamma_5 S_{\rm HTL}(\omega - k_0, \vp-\vk) \biggr].
\label{eq9}
\eea
Here $N_c=3$ is the number of colors, and the sum over $n$ denotes the sum 
over the fermionic Matsubara frequencies $k_0=(2n+1)i \pi T$.

The HTL resummed quark propagator is conveniently expressed by its spectral 
function $\rho_{\rm HTL}$
\bea
&& \hspace*{-0.8cm}S_F(k_0, \vk)= -(\gamma_0 \; k_0 - \vg\cdot \vk)\;
\int_0^{1/T} \;{\rm d}\> \tau \> {\rm e}^{k_0\tau}\nonumber \\
&& \hspace*{-0.8cm}\times \int_{-\infty}^{\infty} {\;{\rm d}\omega}\> 
\rho_{\rm HTL} (\omega, \vk)\; [1-n_F(\omega)]\; {\rm e}^{-\omega \tau}, 
\label{eq10}
\eea
where $n_F(\omega) = 1/[1+\exp(\omega / T)]$ and 
\bea
\rho_{\rm HTL} (k_0, \vk) \hspace*{-0.2cm}& = &\hspace*{-0.2cm} 
{1\over 2}\> \rho_{+} (k_0, k)(\gamma_0 - i\; \hat{\vk}\cdot \vg )\nonumber \\
\hspace*{-0.2cm}&+&\hspace*{-0.2cm}
{1\over 2}\> \rho_{-} (k_0,k)(\gamma_0 + i\; \hat{\vk}\cdot \vg ) 
\label{eq11}
\eea
with \cite{Bra90b}
\begin{eqnarray} 
\rho_{\pm} (k_0, k) \hspace*{-0.2cm}&=&\hspace*{-0.2cm} 
{k_0^2 - k^2 \over 2 m_q^2} 
[\delta (k_0 - \omega_{\pm}) + \delta (k_0 + \omega_{\mp})]\nonumber \\
\hspace*{-0.2cm}&+&\hspace*{-0.2cm}
\beta_{\pm} (k_0, k) \Theta(k^2 -k_0^2).
\label{eq12}
\eea
Here the first part of (\ref{eq12}) corresponds to the pole contribution 
of the HTL propagator. The second part, corresponding to the cut contribution 
from the imaginary part of the HTL quark self energy, is given by 
\bea
\beta_{\pm} (k_0, k) \hspace*{-0.2cm}&=& \hspace*{-0.2cm}
-{m_q^2 \over 2}\> (\pm k_0 - k)\nonumber\\
&& \hspace*{-2.2cm} \times \Biggl \{\biggl[k(-k_0 \pm k) + m_q^2 
\biggl( \pm 1 - {\pm k_0 -k \over 2k}
\ln{k+k_0 \over k-k_0} \biggr)\biggr]^2\nonumber \\
&& \hspace*{-2.2cm} +\biggl[ {\pi \over 2}
m_q^2 {\pm k_0 -k \over k} \biggr]^2 \Biggr \}^{-1}.
\label{eq12a}
\end{eqnarray} 

Combining (\ref{eq8}) to (\ref{eq12a}) the spectral function of the 
pseudo-scalar correlator can be written 
as
\begin{eqnarray}
&& \hspace*{-0.8cm} \sigma_{\rm ps} (\omega,\vp) = 
2N_c ({\rm e}^{\omega/T} -1) \int {{\rm d}^3k \over (2\pi )^3} \nonumber \\ 
&& \hspace*{-0.8cm} \times \int_{-\infty}^{\infty} {\rm d}x {\rm d}x'
n_F(x) n_F(x') \delta(\omega -x-x')\label{eq13} \\
&~&\hspace*{-0.8cm}\times \biggl\{ (1-\vq\cdot \vk)
[\rho_+(x,k) \rho_+(x',q) + \rho_-(x,k) \rho_-(x,q)]\nonumber \\
&~&\hspace*{-0.8cm}+(1+\vq\cdot \vk)
[\rho_+(x,k) \rho_-(x',q) + \rho_-(x,k) \rho_+(x,q)]\biggr\},
\nonumber
\end{eqnarray}
where $\vq =\vp-\vk$. Considering only the pole-pole contribution
originating from the first equation of (\ref{eq12}) all the integrals in
(\ref{eq13}) can be done exactly. One finds
\begin{eqnarray}
&& \hspace*{-0.8cm}\sigma^{\rm pp}_{\rm ps} (\omega , \vp =0) =
{N_c\over 2\pi^2} \frac{({\rm e}^{\omega/T} -1)}{m_q^4}\nonumber \\
&& \hspace*{-0.8cm}\times \biggl[ n_F^2(\omega_+(k_1)) 
\frac{\bigl( \omega_+^2(k_1) -k_1^2 \bigr)^2\, k_1^2}{2| \omega_+'(k_1)|} 
\label{eq14}\\
&&\hspace*{-0.8cm}+2 \sum_{i=1}^2
n_F(\omega_+(k_2^i)) \bigl[1-n_F(\omega_-(k_2^i))\bigr]\nonumber \\
&&\hspace*{-0.8cm}\times \frac{\bigl(\omega_-^2(k_2^i) -(k_2^i)^2 
\bigr)\, \bigl(\omega_+^2(k_2^i) -(k_2^i)^2 \bigr)\,
(k_2^i)^2}{| \omega_+'(k_2^i)-\omega_+'(k_2^i)|}
\nonumber \\
&&\hspace*{-0.8cm}+ \sum_{i=1}^2
n_F^2(\omega_-(k_3^i)) \frac{\bigl(\omega_-^2(k_3^i) -(k_3^i)^2 \bigr)^2 
\, (k_3^i)^2}{2| \omega_-'(k_3^i)|}\; \biggr].  
\nonumber
\end{eqnarray}
Here $\omega_\pm (k)$ denote, as before,
the quark dispersion relations for the ordinary 
quark (+) and the plasmino (-) branch, 
$k_1$ is the solution of $\omega - 2 \omega_+(k_1) = 0$, $k_2^i$ and $k_3^i$
are the solutions of  $\omega - \omega_+(k_2^i)+\omega_-(k_2^i) =0$ and  
$\omega - 2  \omega_-(k_3^i) = 0$, respectively. 
Note that for small momenta the last two
equations can each have two solutions. Furthermore,  
$\omega_\pm'(k)\equiv ({\rm d}  \omega_\pm (x) / {\rm d} x)|_{x=k}$.

Eq. (\ref{eq14}) is shown in Fig.3 for $m_q/T=1$, i.e., $g=\sqrt{6}$. 
The first part of (\ref{eq14}) corresponds to the annihilation of 
collective quark-antiquarks and sets in at the threshold $\omega
\geq 2m_q$. For large energies this is the dominating contribution which
approaches the result obtained from a bare quark propagator (crosses)
for $\omega \gg m_q$. The second part corresponds to a transition from
the upper quark branch to the lower plasmino branch. It starts at $\omega =0$
and terminates with an Van Hove singularity at $\omega =0.47\, m_q$ , where
the difference $\omega_+(k_2^i)-\omega_-(k_2^i)$ has a maximum due to
the minimum in the plasmino dispersion. The third
part describes the annihilation of plasminos and antiplasminos which starts
at the threshold $\omega = 1.86\, m_q$, where another Van Hove singularity
occurs due the minimum of the plasmino branch $\omega_-(k_3^i)$. 
For larger momenta this contribution vanishes quickly due to the exponentially
suppressed spectral strength of the plasmino branch. The singularities
in (\ref{eq14}) can be integrated leading to finite results for
the temporal correlator \cite{Kar00}. Also in Fig.3 the smooth pole-cut and
cut-cut contributions are shown. In particular the latter is suppressed
by about an order of magnitude. 

\begin{figure}
\hspace*{-0.5cm}
\centerline{\psfig{figure=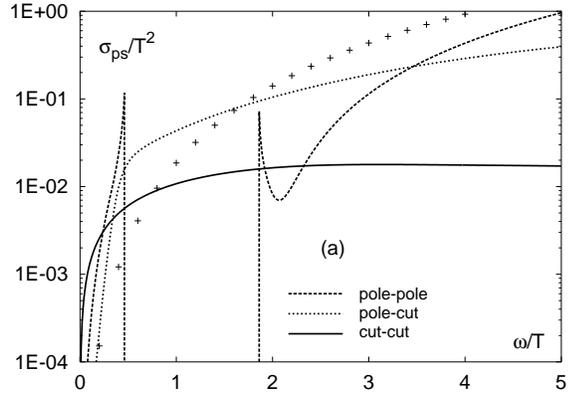,width=8cm}}
\vspace*{-0.5cm}
\caption{Pole-pole, pole-cut, and cut-cut contributions (solid lines)
to the spectral function of the pseudo-scalar temporal correlator at $m_q/T=1$.
The crosses indicate the free spectral function.}
\end{figure}

In the case of the spectral function of the vector correlator
\cite{Kar00} and the dilepton production rate at zero momentum \cite{Bra90b},
for which an effective HTL vertex has to be used, the 
pole-pole and pole-cut contributions are similar (apart from some 
pre-factors). In particular the Van Hove singularities are at the same 
positions. The cut-cut contribution, however, diverges for small
$\omega$ covering up the low-energy Van Hove peak and leading to
an infrared divergent vector correlator.

The temporal correlator follows from its spectral function by an energy 
integration \cite{Kar00}
\bea
G_{ps}(\tau)&=&\int_0^\infty {\rm d}\omega \> \sigma_{ps} (\omega ,\vp=0)
\nonumber \\
&\times & \frac{\cosh(\omega(\tau -\beta /2))}{\sinh(\omega \beta /2)},
\label{eq14a}
\eea
where $\beta = 1/T$ and $\tau $ is restricted to the Euclidean time
interval $[0,\beta ]$. Using (\ref{eq14a}) together with (\ref{eq14})
and the corresponding expressions for the pole-cut and the cut-cut
contributions the temporal pseudo-scalar and vector correlators turn out be 
similar to the free correlation functions containing only bare quarks.
This is caused by a competing effect
of the pole-pole contribution and the pole-cut and cut-cut contributions.
Whereas the pole-pole contribution reduces the spectral function due
to the presence of the effective quark
mass compared to the free spectral function,
there is an enhancement due to the pole-cut and cut-cut contributions
describing higher order diagrams involving external gluons. Surprisingly
the two effects compensate each other in the temporal correlator
almost completely. Hence the deviation from the free correlator, observed
on the lattice, cannot be attributed to HTL medium effects \cite{Kar00}.

\section{Gluon condensate quark propagator}

The problem connected with the above calculations relies in its perturbative 
nature. Since the coupling constant $g$ is not small in realistic situations,
the extrapolation to relativistic heavy ion collisions is questionable.
Furthermore due to collinear and magnetic divergences the convergence
even of the HTL improved
perturbative series at finite temperature is upset at least for
low mass $M\simeq g^2T$ dileptons \cite{Aur00}.

In Ref.\cite{Sch99} another effective quark propagator has been 
constructed by taking into account the gluon condensate measured
in lattice QCD calculations above the phase transition. In this way
non-perturbative effects are included, which allow to study realistic
temperatures addressed on the lattice. For this purpose we calculate 
the quark self energy from a one-loop diagram using a non-perturbative
gluon propagator that contains the gluon condensate analogously to
the zero temperature case \cite{Lav88}.
Of course, the approach is a purely phenomenological 
combination of Green functions with lattice results.
The effective quark propagator can then be written as 
in (\ref{eq6}) and (\ref{eq7}) with
\begin{eqnarray}
a &=& - \frac{g^2}{6} \frac {1}{K^6} \biggl [ \left ( \frac {1}{3} k^2
- \frac{5}{3}k_0^2\right )
\langle {\mathcal E}^2 \rangle_T\nonumber \\
&& - \left ( \frac{1}{5}k^2
- k_0^2 \right ) \langle {\mathcal B}^2 \rangle_T \biggr ] , \nonumber \\
b &=& - \frac{4}{9}g^2 \frac {k_0}{K^6} \left [ k_0^2\langle {\mathcal E}^2 
\rangle_T 
+ \frac{1}{5}k^2\langle {\mathcal B}^2 \rangle_T \right ] ,  
\label{eq15}
\end{eqnarray}
where the in-medium chromoelectric, $\langle {\mathcal E}^2\rangle_T$, and 
chromomagnetic condensates, $\langle {\mathcal B}^2\rangle_T$, are taken from
lattice calculations \cite{Boy96}.

In Fig.4 the dispersion relations following from the poles of this
propagators are shown at $T=2T_c$ \cite{Sch99}. 
The important point here is that, 
although the HTL resummed and the gluon condensate quark propagators
are completely different, the dispersion relations show the very same
behaviour. In particular we observe again the minimum in the plasmino branch.
Therefore we expect again Van Hove singularities in the dilepton rate
following from this quark propagator. In a first step we neglected 
an effective quark-photon vertex, related to the gluon condensate quark
propagator by Ward identities \cite{Mus00}. However, as we discussed above,
this will not change the existence and positions of Van Hove singularities
following solely from the quark dispersion relations. In Fig. 5 the dilepton
production rate at zero photon momentum versus the invariant photon mass $M$ 
from this investigation \cite{Mus00a} is shown for
different temperatures. The interpretation of the various curves is identical 
with the one in the last section. However, in this case no pole-cut and cut-cut
contribution are present as the quark self energy containing the gluon
condensate has no imaginary part \cite{Sch99}.

\section{General quark dispersion relations}

\begin{figure}
\vspace*{-0.7cm}
\hspace*{-0.5cm}
\centerline{\psfig{figure=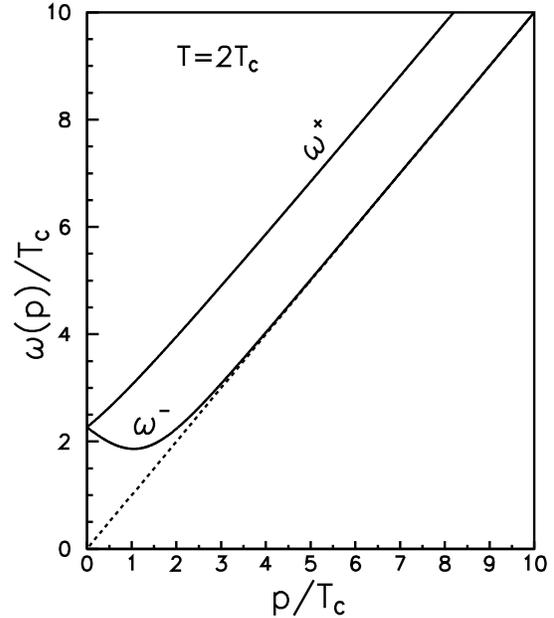,width=9.3cm}}
\vspace*{-3cm}
\caption{Quark dispersion relations according to the gluon condensate quark
propagator at $T=2T_c$.}
\end{figure}

The two completely different approximations for the quark propagator,
discussed in the last two sections, lead to the same qualitative behaviour 
of the dispersion relations. Hence one may speculate that the
dispersion relations shown in Fig.1 and Fig.4 are a consequence of
the most general full propagator for massless fermions at finite temperature.
Analyzing the most general expression for the in-medium fermion propagator
(\ref{eq6}) and (\ref{eq7}) one can show the following general
properties of the dispersion relations \cite{Pes00}: (1) there are always two
branches; (2) both branches start at the same energy, i.e. effective
fermion mass, at zero momentum; (3) the both branches have an opposite slope 
at zero momentum; (4) the branch corresponding to the usual fermion 
approaches the free dispersion relation, $\omega =k$, for large momenta $k$.
Assuming that the plasmino branch also approaches the free dispersion
at large momenta, as it does in the two independent examples discussed above,
the plasmino branch must always possess a minimum. So far this assumption has
not been proven. Maybe future lattice calculations might be capable of
investigating the quark dispersion relations \cite{Kar99}. However, it 
appears to be reasonable that the plasmino minimum is a general feature
leading to Van Hove singularities in the dilepton production rate and
mesonic temporal correlators.

\section{Consequences for relativistic heavy ion collisions}

\begin{figure}[t]
\vspace*{-0.7cm}
\hspace*{-0.2cm}
\centerline{\psfig{figure=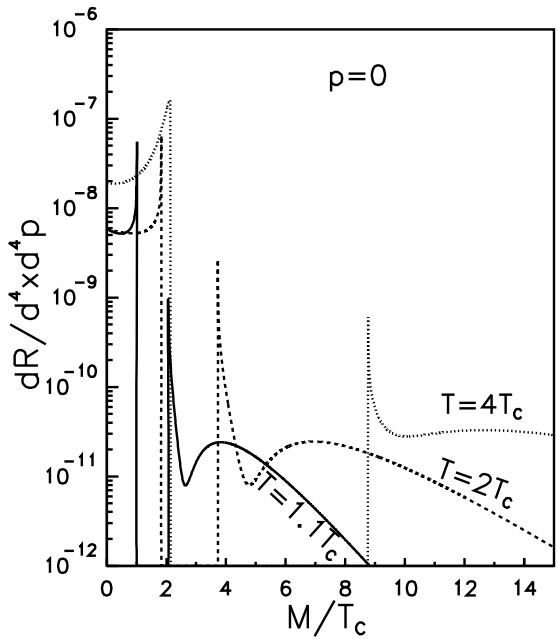,width=9.3cm}}
\vspace*{-3cm}
\caption{Dilepton production rate at zero photon momentum versus the 
invariant photon mass $M$ following from the gluon condensate quark 
propagator.}
\vspace*{-0.7cm}
\end{figure}

Now the important question is whether the Van Hove singularities 
in the QGP can be observed in relativistic heavy ion collisions.
One possibility would be to look at the low-mass dilepton spectrum,
which has been investigated already at SPS energies \cite{Cer98}.
Here no sign for interesting structures due to Van Hove singularities
has been found. However, hydrodynamic simulations \cite{Sol97}
show that the QGP contribution, if there is a QGP phase at all 
as indicated at SPS \cite{Hei00}, are suppressed at least by an order of
magnitude compared to the hadronic contributions. This situation
will change at RHIC and LHC, where the QGP contribution should
dominate the thermal emission from the fireball. If low-mass
dileptons should be investigated at RHIC or LHC, interesting structures
coming from the Van Hove peaks and the gap in the dilepton rate
might appear. From a comparison of the equation of state
found in lattice calculations and quasiparticle models \cite{Pes94}
effective quark masses of the order 500 MeV are expected.
Therefore the structures may show up at invariant masses below
about 1 GeV.

To what extent these distinct structures will survive in heavy ion collisions
depends on how much of it will be covered and smeared out
by higher order effects, e.g. bremsstrahlung and damping. Also finite
momenta \cite{Mus00a,Won92} and the space-time evolution of the fireball
will wash out the structures to some degree. These effects are an 
interesting subject for future investigations. Anyway, if a non-trivial
structure should be observed in the low-mass dilepton spectrum it would
provide an unique signal for the QGP formation,
since one does not expect a similar signal from the hadronic phase. 
Due to in-medium collisions between hadrons the structures from
the hadronic phase such as the $\rho$-peak are washed out \cite{Rap99}.
Hence structures in the low mass dilepton spectrum due to Van Hove 
singularities would not only indicate a deconfined phase, but
also the existence of collective quark modes in the QGP.

\section{Conclusions}

The quark dispersion relations in the QGP have been studied using two 
different approximations for the quark propagator. First the HTL 
method, which is based on the resummation of the perturbative quark self
energy obtained in the high temperature limit has been adopted. Secondly
the gluon condensate measured on the lattice above the critical
temperature has been included in the effective quark propagator.
The poles of the effective quark propagators, following from these
completely different approaches, determine quark dispersion relations
which show the same behaviour. In both approaches two branches, of which 
one shows a minimum, namely the plasmino branch, have been found. There 
are strong indications that this is a general feature of massless 
fermions at finite temperature. 

The minimum in the plasmino branch leads to a vanishing group velocity of
the plasma modes and therefore to a diverging density of states in 
the low-mass dilepton production rate and in the spectral function
of mesonic correlators. Therefore Van Hove singularities appear
in these quantities. The dilepton production might serve as a promising 
signature for the QGP formation in relativistic heavy ion collisions.
Whether these Van Hove singularities can be observed in the dilepton spectrum
from a dynamical QGP,
produced possibly in the fireball of a nucleus-nucleus collision, is
an open question. Anyway it will be worthwhile to look for non-trivial
structures, indicating the existence of deconfined, collective quark modes, 
in the low-mass dilepton spectrum at RHC and LHC. 

The meson correlation 
functions, on the other hand, can be compared to lattice calculations, 
which allows to extract informations on the non-perturbative nature of 
the QGP. Future QCD lattice calculations might be able to investigate
directly the quark dispersion relations and the spectral functions of 
the mesonic correlators \cite{Wet00}, 
which contain much more informations than the 
correlators themselves. Summarizing, the verification of Van
Hove singularities in the QGP experimentally as well as on the lattice
will be an interesting challenge.

\vspace*{0.5cm}

\centerline{\bf ACKNOWLEDGMENTS}
\vspace*{0.3cm}
Most of the results presented here have been obtained in collaboration
with F. Karsch, M. Mustafa, A. Peshier, and A. Sch\"afer.

\end{document}